\documentclass[html]{article}
\usepackage{epsfig}
\usepackage{graphicx}
\usepackage{amssymb}

\begin{document}
\title{Numerical Simulation of the Perrin - like Experiments}
\author{Zygmunt Mazur \\
Institute of Experimental  Physics, Wroc{\l}aw University, pl. M.
Borna 9,\\ 50-405 Wroc{\l}aw, Poland
\\  and \\
  Dariusz Grech \\
Institute of Theoretical Physics, Wroc{\l}aw University, pl. M.
Borna 9, \\ 50-405 Wroc{\l}aw, Poland }

 \maketitle

 \begin{abstract} A simple model of random Brownian walk of a spherical
mesoscopic particle in viscous liquids is proposed. The model can
be both solved analytically and simulated numerically. The
analytic solution gives the known Einstein-Smoluchowski diffusion
law $\langle r^2\rangle = Dt$ where the diffusion constant $D$ is
expressed by the mass and geometry of a particle, the viscosity of
a liquid and the average effective time between consecutive
collisions of the tracked  particle with liquid molecules. The
latter allows to make a simulation of the Perrin experiment and
verify in detailed study the influence of the statistics on the
expected theoretical results. To avoid the problem of small
statistics causing departures from the diffusion  law we introduce
in the second part of the paper the idea of so called Artificially
Increased Statistics (AIS) and prove that within this method of
experimental data analysis one can confirm the diffusion law and
get a good prediction for the diffusion constant even if
trajectories of just few particles immersed in a liquid are
considered.
\end{abstract}

\section{Introduction}
Recently a big progress has been made in application of digital
technique in experimental physics what allows to perform milestone
physics experiments even in student laboratories. A good example
is the Perrin experiment [1] considered as the first one directly
proving the atomic structure of matter. However, its verification
at university laboratories [2],[3],[4],[5] due to small statistics
one takes, may meet some difficulties (see e.g.[2],[4],[5]). The
linear dependence between the average square displacement $\langle
r^{2}\rangle$ of the particle in media due to its Brownian motion
and the observation time $t$ as
required by the Einstein-Smoluchowski diffusion law often becomes very problematic.\\
It is essential therefore to examine the minimal statistics
(number of tracked particles) one should take into account in the
limited observation time to reveal the major feature of the
diffusion law. We propose the analytical model which can be also
easy simulated numerically. The aim of this model is to
investigate how the results of $\langle r^{2}\rangle$ versus $t$
depend on the statistics and what the scaling range of the
expected linear relationship is. This study should help to set up
the experiment properly as well as to analyze  the
obtained results more correctly.\\
In the next section we present the model directly reflecting the
physics standing behind the Perrin experiment. In section 3 the
results of numerical simulation of this model are described and
main features of diffusion relation with its scaling range are
revealed for various number of particles to be observed. To avoid
the problem of small statistics causing departures from the strict
power law behavior we introduce and discuss the idea of
Artificially Increased Statistics(AIS) in section 4. This method
is then applied both to the results of numerical simulation and to
some experimental data. We argue the method may significantly
decrease the level of statistical noise in data, leading to much
better agreement with the linear dependence in diffusion theory.
In the last section summary of obtained results is given.

\section{Description of the Model}
The most popular derivation of the diffusion law in media with
viscosity comes from Einstein [6], Langevin [7] and Smoluchowski
[8]. Here we propose another approach based on the time series
analysis combined with the average time $\tau$ between consecutive
collisions of the tracked mesoscopic particles with other
particles in liquid (i.e. $\tau$ has the meaning of the average
time between collisions which significantly change the motion of
the tracked object). Such approach seems to be closer to the
spirit of the original Perrin experiment [1].
\\
Let the trajectory of the observed  particle of mass $m$ moving in
$d$-dimensional space is $x^{\alpha}(t)$, where $\alpha
=1,2,...,d$. We assume $x^{\alpha}(t)$ to be discrete
$d$-dimensional time series with constant spacing $\tau$ in time
($t = 0, \tau, 2\tau,...,N\tau$). The obvious notation
\begin{equation}\label{eq1}
x^{\alpha}(k\tau) = x^{\alpha}_{k}, k = 1, 2, ..., N
\end{equation}

and

\begin{equation}\label{eq2}
\Delta x^{\alpha}_{k} = x^{\alpha}_{k+1}-x^{\alpha}_{k}
\end{equation}
will be applied, where $\Delta x^{\alpha}_{k}$ is the instantenous
displacement of the particle at $t=k\tau$.

For the stationary, integer Brownian motion
(no displacement correlation)
 with no drift one has for large $n$:
\begin{equation}\label{eq3}
\langle\Delta x^{\alpha}_{i}\rangle_{n} = 0
\end{equation}

and

\begin{equation}\label{eq4}
\langle\Delta x^{\alpha}_{i}\Delta x^{\alpha}_{j}\rangle_{n} =
\delta_{ij}{({\sigma}^{\alpha}_{i})}^2
\end{equation}
where $\langle .\rangle_n$ is the average taken over the ensamble
of $n$ tracked particles and ${({\sigma}
^{\alpha}_{i})}^2=\sigma^{2}$ is the standard deviation of
instantenous displacements, i.e.:

\begin{equation}\label{eq5}
\langle (\Delta x^{\alpha}_{i})^{2}\rangle_{n} = \sigma^{2}
\end{equation}

The total mean squared displacement $\langle r^{2}\rangle_{n}$ of
particles from their initial positions after $N$ collisions can be
easy calculated with the help of Eq.~(5):

\begin{equation}\label{eq6}
\langle \Delta r^{2}\rangle_{n} =
\langle\sum\limits^{d}_{\alpha=1}(\sum^{N}_{i}\Delta
x^{\alpha}_{i})^{2}\rangle_{n} = \frac{d\sigma^{2}}{\tau}t
\end{equation}

In order to calculate $\sigma^{2}$ let us notice that

\begin{equation}\label{eq7}
\Delta x^{\alpha}_{i} = \tau\langle v^{\alpha}_{i}\rangle_{\tau}
\end{equation}
with $\langle v^{\alpha}_{i}\rangle_{\tau}$ being the average
velocity of the $i$ - th particle between collisions. Hence from
Eqs.~(5) and(7):

\begin{equation}\label{eq8}
\sigma^{2} = \tau^{2}\langle\langle
v^{\alpha}_{i}\rangle^{2}_{\tau}\rangle_{n}
\end{equation}

The equipartition theorem establishes the connection of
microscopic quantities with the absolute temperature $T$
and the Boltzmann constant $k$:

\begin{equation}\label{eq9}
\frac{1}{2}m\langle\langle
v^{\alpha}_{i}\rangle^{2}_{\tau}\rangle_{n} = \frac{1}{2}kT
\end{equation}

Therefore Eq.~(6) reads:

\begin{equation}\label{eq10}
\langle\Delta r^{2}\rangle_{n} = (\frac{dkT}{m}\tau)t
\end{equation}

The above formula is the standard diffusion law with the diffusion
constant

\begin{equation}\label{eqx11}
D = \frac{dkT}{m}\tau
\end{equation}
expressed in terms of $\tau$.\\

Usually one writes $D$ in terms of liquid viscosity $\eta$ as

\begin{equation}\label{eqx12}
D = \frac{dkT}{\alpha}
\end{equation}
where $\alpha = 6\pi \varrho \eta$ (Stokes law) and $\varrho$
being the radius
of the considered mesoscopic particles.\\
Hence one gets the simple relation between parameter $\tau$ in the
model and macroscopic quantities $m$, $\alpha$:

\begin{equation}\label{eq13}
\tau = \frac{m}{\alpha}
\end{equation}

Thus the model reproducing the known diffusion law also estimates
the average time $\tau$ lapsing between consecutive collisions in
the system as the simple function of macroscopically measured
quantities. This time can be taken as the input parameter in the
numerical study of the Perrin experiment what is done in the next
section.

\section{Numerical Simulation of the Perrin Experiment}
The solution in Eq.~(10) can be checked via numerical simulation of
the Brownian motion in viscous media. In fact this simulation is
the only way one can find the sufficient statistics, i.e. the
number of tracked particles in the ensamble one should observe in
real experiment to obtain results confirming the linear relation.
If sufficient statistics requirement is not satisfied, one
observes significant departures from the linear behavior $\langle
r^2\rangle_{n}\sim t$ (see e.g. Ref. [2], [4]).\\
We simulated all time series $\{x^{\alpha}_i\}$ in $d=2$
dimensions usually discussed by experimentalists. The time series
were built in the well known iterative way

\begin{equation}\label{eqx14}
x^{\alpha}_{i+1}= x^{\alpha}_{i} + \Delta x^{\alpha}_{i}
\end{equation}

\begin{equation}\label{eq15}
r^{2}_{i} = {(x^{1}_{i})}^{2} + {(x^{2}_{i})}^{2}
\end{equation}
where displacements  have been generated as the random gaussian
numbers $N(0,\sigma)$, with the standard deviation $\sigma =
\tau{(kT/m)}^{1/2}$ obtained from Eqs.~(8),(9). All simulations
were performed for the case of diffusion in pure water $(\eta =
1.00\times {10}^{-3} Pa\cdot s)$, room temperature $T = 295~K$, $m
= 4.28\times {10}^{-16}~kg$ and $\varrho = 425~nm$
what roughly corresponds to the real Perrin experiment parameters.\\
The essential task to be done just in the beginning was to
determine the scaling range $\lambda$ of the discussed linear
dependence as a function of the number of tracked particles $n$.
It was done for the bunch of simulated trajectories  varying the
number of observed particles in the range $n = 10 \div 500$. The
bunch of twenty trajectories was investigated for any $n$ in the
above range. The examples of just five runs in each bunch (for the
clarity of figure we do not show all the runs) are pictured in
Fig.1 a-d. Hence we have found the scaling range relation revealed
in Fig.2. The best fit gives
\begin{equation}\label{eqx16}
\lambda \sim n^\beta
\end{equation}
where $\beta = 0.51 \pm 0.04$ and the uncertainty comes
from the statistics.\\
Let us notice that if the number of observed particles does not
exceed $10$ the linear dependence $\langle r^2\rangle_{n}\sim t$
can be confirmed only  for the observation time $t<3s$! It makes
the analysis taking into account longer observation times (as
authors
of Ref.~[4] did), simply incorrect.

Having the scaling range determined we may proceed to calculate
the diffusion constant value and its expected standard deviation
from the mean. Such analysis was done by us for the simulated
trajectories mentioned above. Some chosen cases (again for the
clarity of graph we do not show all of them) with maximal and
minimal values of $D$ for every $n$ are shown in Fig.3 a-d. All
 results of the mean $D$ values and their standard deviation as
the function of $n$ are presented in Fig.4.

Hence we see that the final result within $10\%$ percent of the
expected theoretical value can be found only if one considers the
ensamble of $n \geq 50$ particles.

\section{Analysis of Results with Artificially Increased Statistics}

The results of the previous section
seem to suggest that
to get a
reasonable agreement with the diffusion law predictions
one should take into account in the real
experiment data from at least $ n\sim 50$ particles.
 In many
less professional labs (e.g. student labs) such a requirement is
virtually impossible to be satisfied - mainly because of the
limited time duration of the data collection if no sophisticated
computerized apparatus is used. Below we give the idea that helps
to overcome such a difficulty. We call it Artificially
Increased Statistics (AIS).\\
The main idea of AIS is to build the statistics of consecutive
displacements from the very small number of available
trajectories, counting all the displacements not from the initial
starting point $(x^1_0, x^2_0) = (0, 0)$ but varying it along the
whole one particle trajectory. Thus any momentary position of the
particle, say $(x^1_k, x^2_k), k=1,2,...,N$ is the starting point
to collect statistics of all displacements afterwards, i.e.
$(\Delta x^1_{l-k}, \Delta x^2_{l-k}), l>k$, where $\Delta
x^{\alpha}_{l-k} = x^{\alpha}_l - x^{\alpha}_k$ is the $\alpha$-th
part of the $(l-k)$ step displacement. This way for the time
series of length $N$ we have $N-m$ data for $m$-steps
displacements instead of just one displacement usually taken into
account. Then the statistics is averaged in the usual way over the
all considered (observed) particles. This way even if $n$ is small
the overall number of data entering  the statistics is large
enough to fulfil the linear law expectation.

Let us now look at the results of the application of AIS to the
simulated Brownian motion as well as to the pure experimental data
from the real experiments.

In Fig.5a-b we present the bunch of squared displacements in time
taken for the statistics of $n=10$(a) and $n=50$(b) particles
worked out with the AIS procedure. The comparison with the "naked"
data from Fig.3a-b shows the tremendous difference. Although the
scaling range after AIS lifting does not seem to  change a lot,
the linear dependence $<r^2> \sim t$ is now much more convincing.
In fact the comparison of diffusion constants $D$ obtained from
the "naked" analysis and from the data lifted by AIS shows about 7
times smaller uncertainty in $D$
\begin{table}[ht]
\caption{The comparison analysis of diffusion constant
 values found as the best fit before and after the AIS procedure.
  Each item is taken from different simulation of the
   n = 10 or n = 50 particles run.
   The scaling range is fixed according to
   Fig 1a,b with the sampling time interval  $\Delta \, t = 10^{5}
   \tau  \sim  0.01 s$.}
   \begin{center}
\begin{tabular}{|c|c|c|c|c|}
\hline
 & \multicolumn{4}{c|}{$D(\mu\, m^2 \, s^{-1} )$}
 \cr
\cline{2-5}
 & \multicolumn{2}{c|}{n=10 particles}  & \multicolumn{2}{c|}{n=50 particles}
\cr
\hline
\hline
Run nr  &   Before AIS  &   AIS data   & Before AIS  &  AIS data
\cr
   &    (``naked`` data)  && (``naked`` data) &
\cr
&&&&
\cr
\hline
1. &     1.04 &  1.83   &   1.92   &   1.92 \cr
\hline
 2.    &    2.07  & 1.76  &  1.90  &  1.95
\cr
\hline
 3.   &     1.28   &   1.90   &   1.88   &   1.95
\cr
\hline
 4.    &    1.65  & 2.15  & 2.29  & 1.83
\cr
\hline
5.
   &     3.23  &    2.00  &    2.24  &    2.16
   \cr
\hline
    6.  & 0.96
&  1.92 &  2.00 &  2.03
\cr
\hline
7. &       2.73 &     1.95 & 1.73   &
2.05
\cr
\hline
 8. &  4.01  &    2.23  & 1.74 &  1.86
 \cr
\hline
 9.  & 1.70   &
1.82   &   2.64  &    2.00
\cr
\hline
 10.  &     1.46  &    1.99 &  1.99 &
1.89
\cr
\hline
\hline
 $< D >$   &          2.00 &     1.96  & 2.03  & 1.96
 \cr
\hline
  $\sigma_{D}$
&  1.0  &     0.15  & 0.28 &  0.10 \cr
\hline
\end{tabular}
\end{center}
\end{table}
evaluation in the case of $n=10$ statistics (see Table 1). The corresponding result for the
$n=50$ case is improved about $3$ times.\\
We have calculated also the mean absolute error (MAE)defined as

\begin{equation}
\delta_{MAE}=\frac{1}{N}\sum\limits^{N}_{k=1}|D_k - D_{th}|
\end{equation}
where $D_{th}$ is the theoretical value of the diffusion
coefficient ($D_{th}=2.01 \mu m^2/s$ for the considered diffusion
process) and the sum is taken over all simulated runs.\\
For the sample of $10$ runs with and without AIS one obtains for
$n=10$ particles $\delta_{MAE}(n=10)=0.80~\mu m^2s^{-1}$
decreasing to $\delta^{AIS}_{MAE}(n=10)=0.13~\mu m^2s^{-1}$ when
AIS is switched on. The corresponding result for $n=50$ particles
are (in $\mu m^2s^{-1}$) $\delta_{MAE}(n=50)=0.20$ and
$\delta^{AIS}_{MAE}(n=50)=0.09$
respectively.\\

The positive feature of AIS procedure can also be seen directly
with the pure experimental data. We show in Fig.~6a the data taken
in Ref.~[4] for the diffusion of $n=5$ latex spherical particles in
the pure water. One gets much better correspondence with the
linear dependence when AIS procedure is applied to these
experimental points, what is clearly revealed in Fig.6b. The
obtained best fit for the diffusion constant corresponds now
closely to the expected theoretical value $D_{th}=2.01
~\mu m^2s^{-1}$ what is not the case of the fit obtained
by authors of Ref.~[4].

\section{Conclusions}
The proper determination of the scaling range for the linear
dependence $\langle r^2\rangle \sim t$ is crucial in  the data
analysis. We argued this scaling range behaves like $\lambda \sim
n^\beta$, where the constant $\beta$ was determined as $\beta \sim
0.5$. The numerical simulation shows that for the case of
mesoscopic particles diffusing in water the scaling range for $n
\sim 10$ particles is as short as $\lambda \leq 3s$. For $n<10$
this scaling range is difficult to determine at all. In many
papers this fact
is ignored what gives misleading results.\\
However, even if one remains in the scaling range regime the
results of simulated runs are not always statistically repeatable
if too small statistics is considered. The minimal number of
tracked particles to reveal the diffusion law is $n\geq 50$. One
may nevertheless find the reasonable correspondence  between
theoretical predictions and experimental results even for the
smaller number of tracked particles if the idea of AIS is applied.
In this paper we have described this idea and have shown how it
works for simulated data as well as for data taken from the real
experiments. It turns out that with AIS analysis one may get
results within $10\%$ of the expected theoretical value of the
diffusion constant tracking just few mesoscopic objects. The
corresponding input data without AIS gives much bigger uncertainty
of the order of $50\%$ (see Fig.4). The same applies when MAE is
calculated. To decrease the uncertainty to the former
level of $10\%$ one has to track roughly ten times more objects!\\
We have checked that for $n=50$ tracked particles AIS procedure
decreases the statistical uncertainty in $D$ from about $15\%$
(the "naked" data case) to $\sim 5\%$. Simultaneously the
$\delta_{MAE}$ drops down about twice (from $0.20~\mu m^2s^{-1}$ to
$0.09~ \mu m^2s^{-1}$). The AIS procedure is here less impressive
than for $n=10$ case but still shows the
significant improvement in data results.\\
This way it is quite possible to collect data giving very good
prediction for the diffusion constant even in less professional
labs where one is not able to measure simultaneously signals
coming from bigger number of objects. Hence, other important
physical constants (like e.g. Boltzmann constant $k$ or the
Avogadro number $N_A$) can be deduced with high accuracy what is
often the crucial point of
such experiments.\\
The simulations  were done by us also for liquids with other
viscosities. The same final conclusions as for the case of water
can be formulated. Because of very similar results we did not show
them explicitly in this paper but we believe they should be
studied in the way of numerical simulation in any case before the
actual experiment is planned.

\begin{figure}[tbh]
    \begin{center}
        \includegraphics[height=5truecm,angle=0]{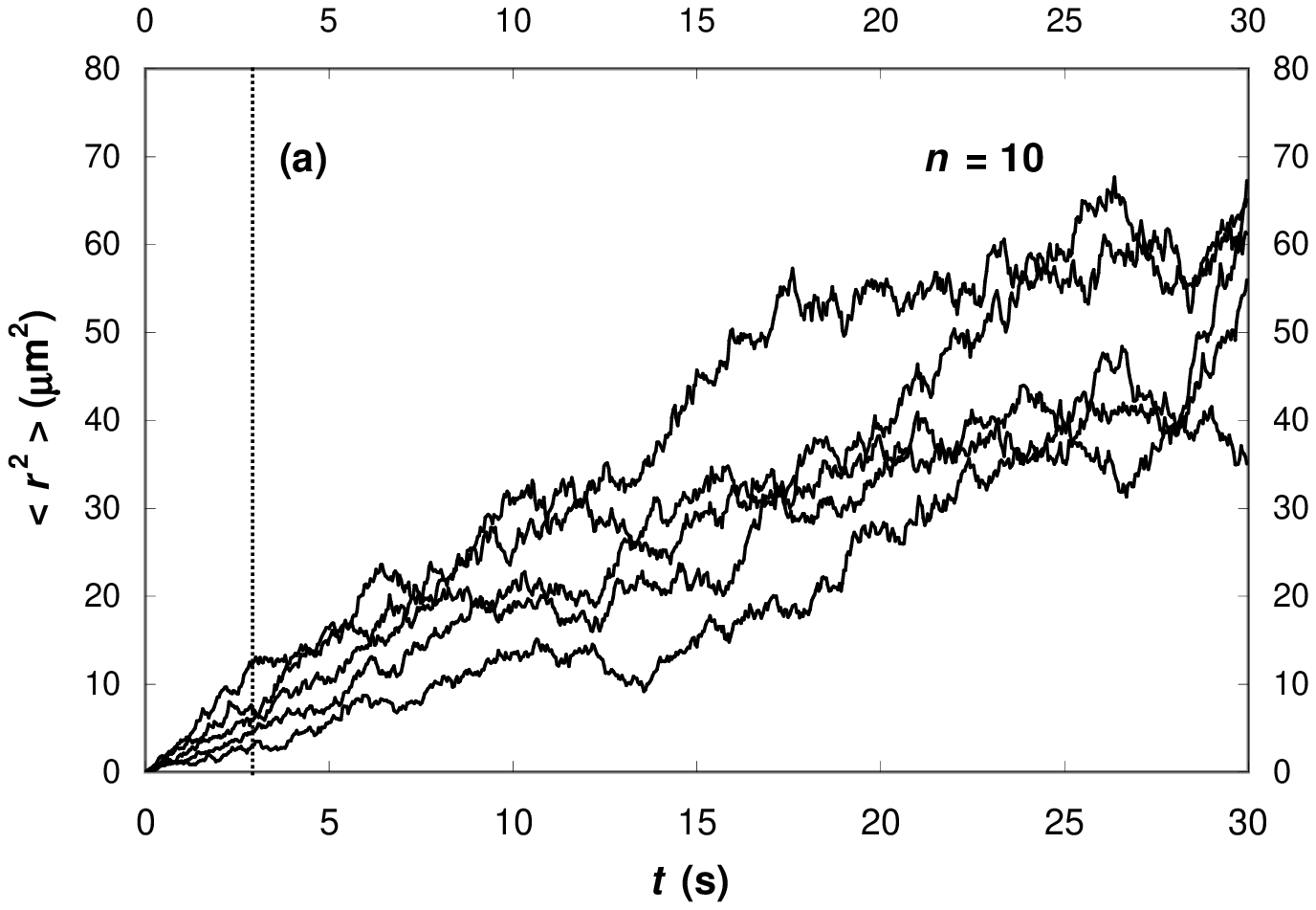}
         \\[4pt]
        \includegraphics[height=5truecm,angle=0]{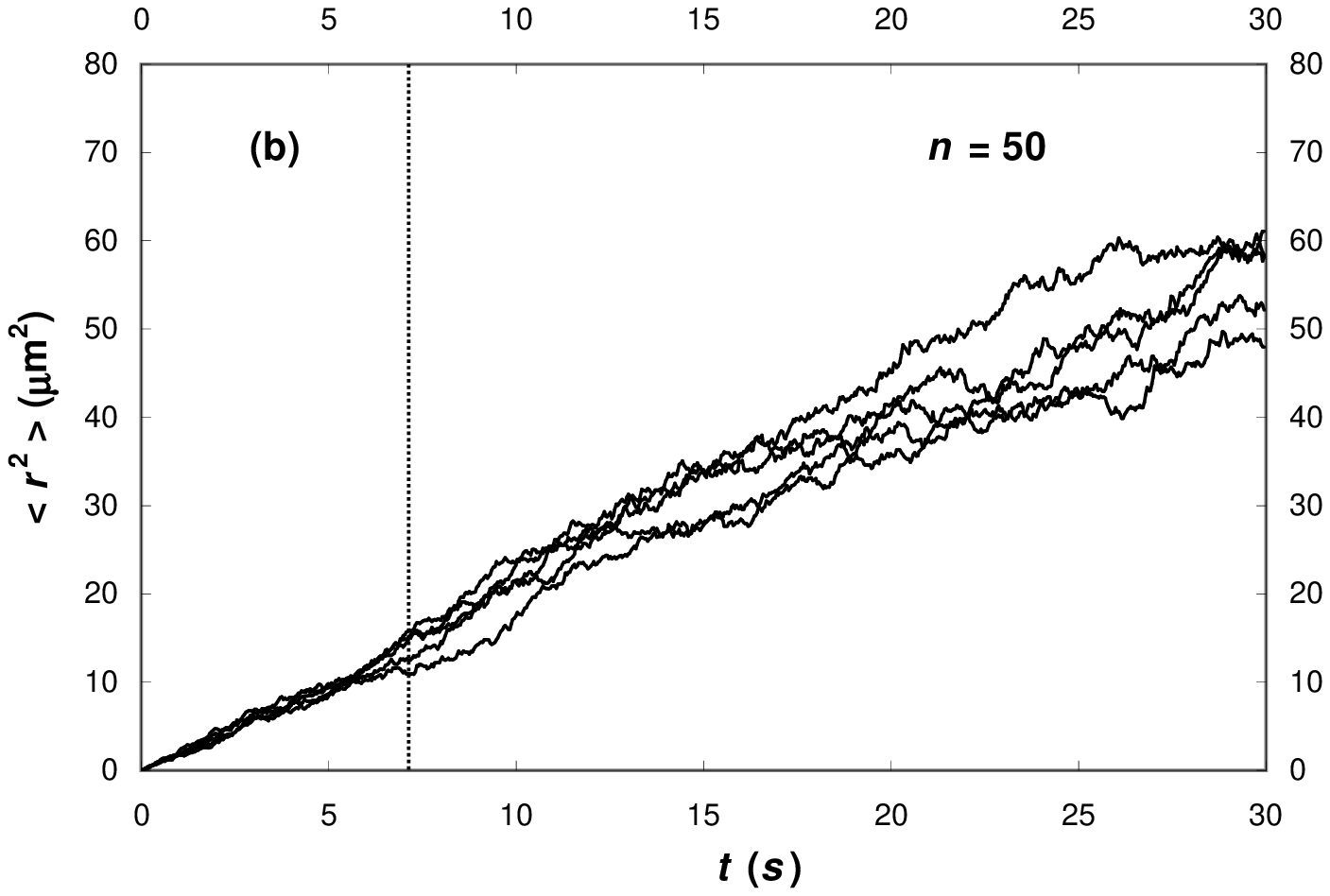}
        \\[4pt]
        \includegraphics[height=5truecm,angle=0]{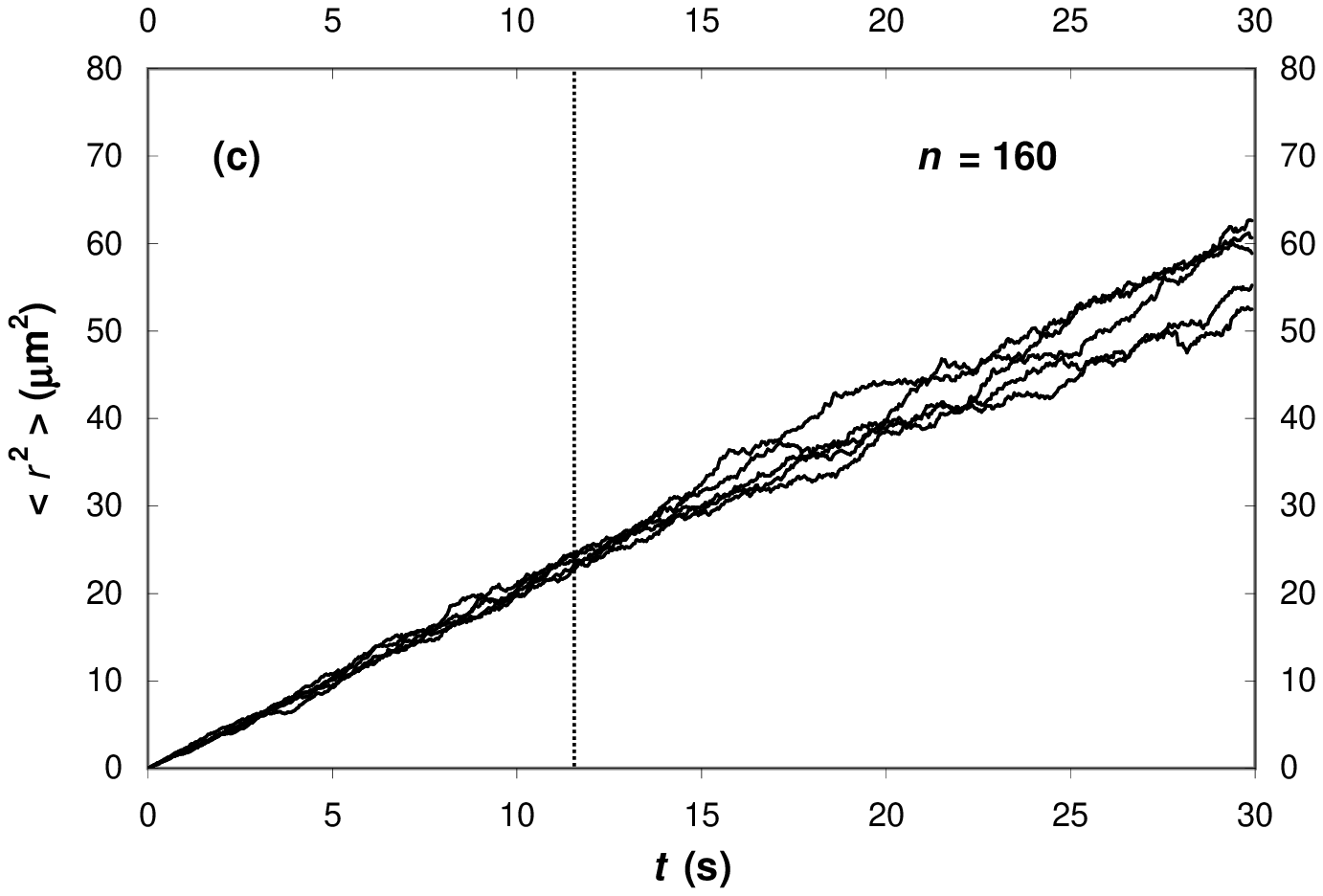}
        \\[6pt]
        \includegraphics[height=5truecm,angle=0]{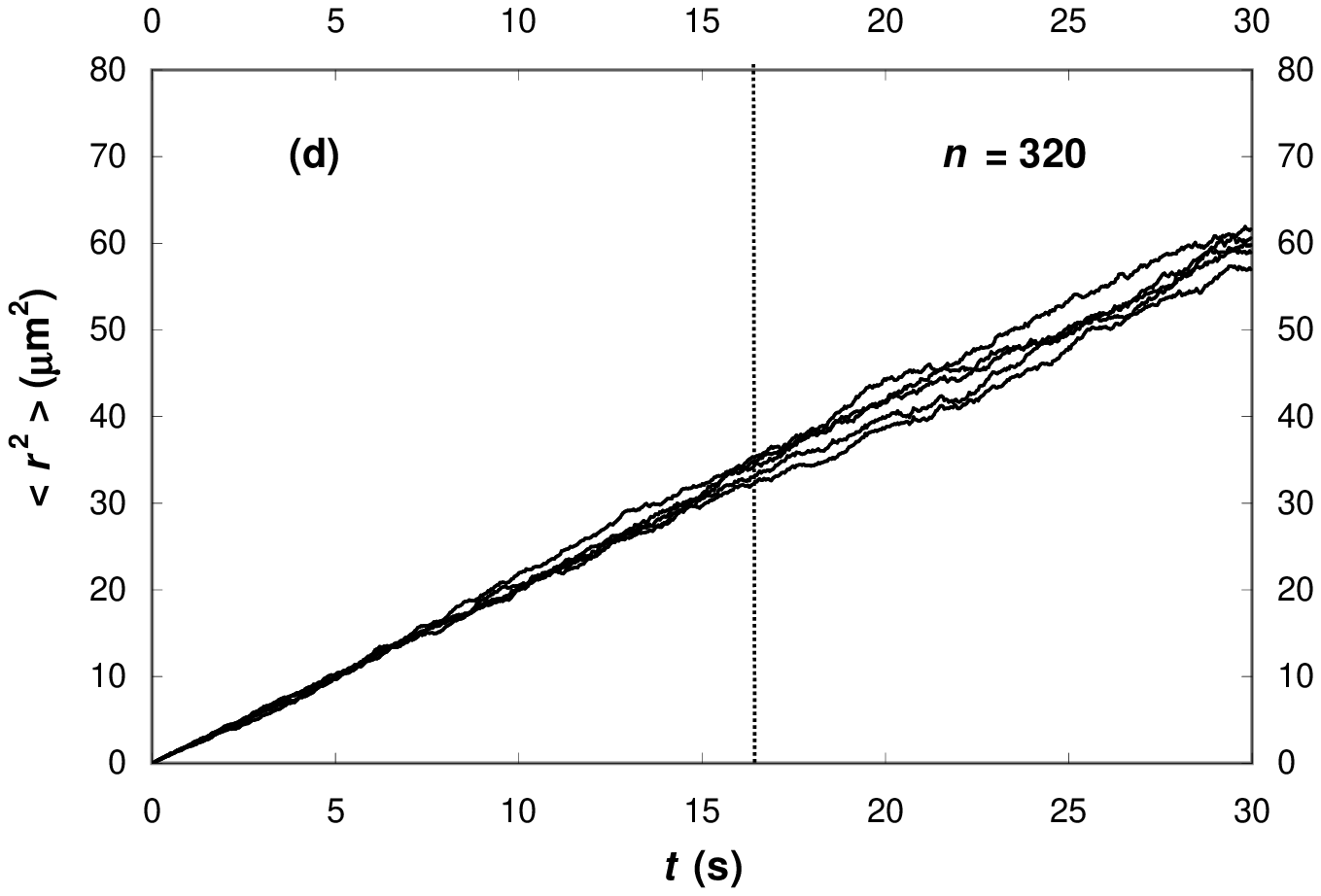}
    \end{center}
    \label{figdgzm1}
\caption{Examples of $\langle r^2\rangle$ dependence versus $t$
for the simulated runs of $n=10$ (a), $n=50$ (b), $n=160$ (c) and
$n=320$ (d) particles. The scaling range of the linear dependence
$\langle r^2\rangle \sim t$ is marked as the vertical line in each
case. }
\end{figure}

\begin{figure}[tbh]
    \begin{center}
       \includegraphics[height=5truecm,angle=0]{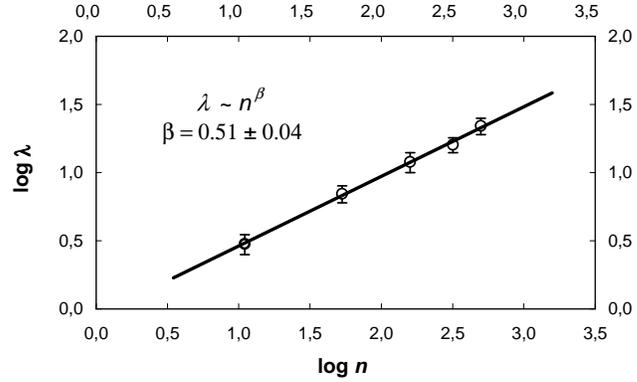}
\caption{The scaling range dependence on the number of tracked
particles $n$. The error bars correspond to statistical
uncertainties in determination of $\lambda$ for given $n$. All
data come from the numerical simulations.}
 \end{center}
  \label{figdgzm2}
\end{figure}

\begin{figure}[tbh]
    \begin{center}
        \includegraphics[height=5truecm,angle=0]{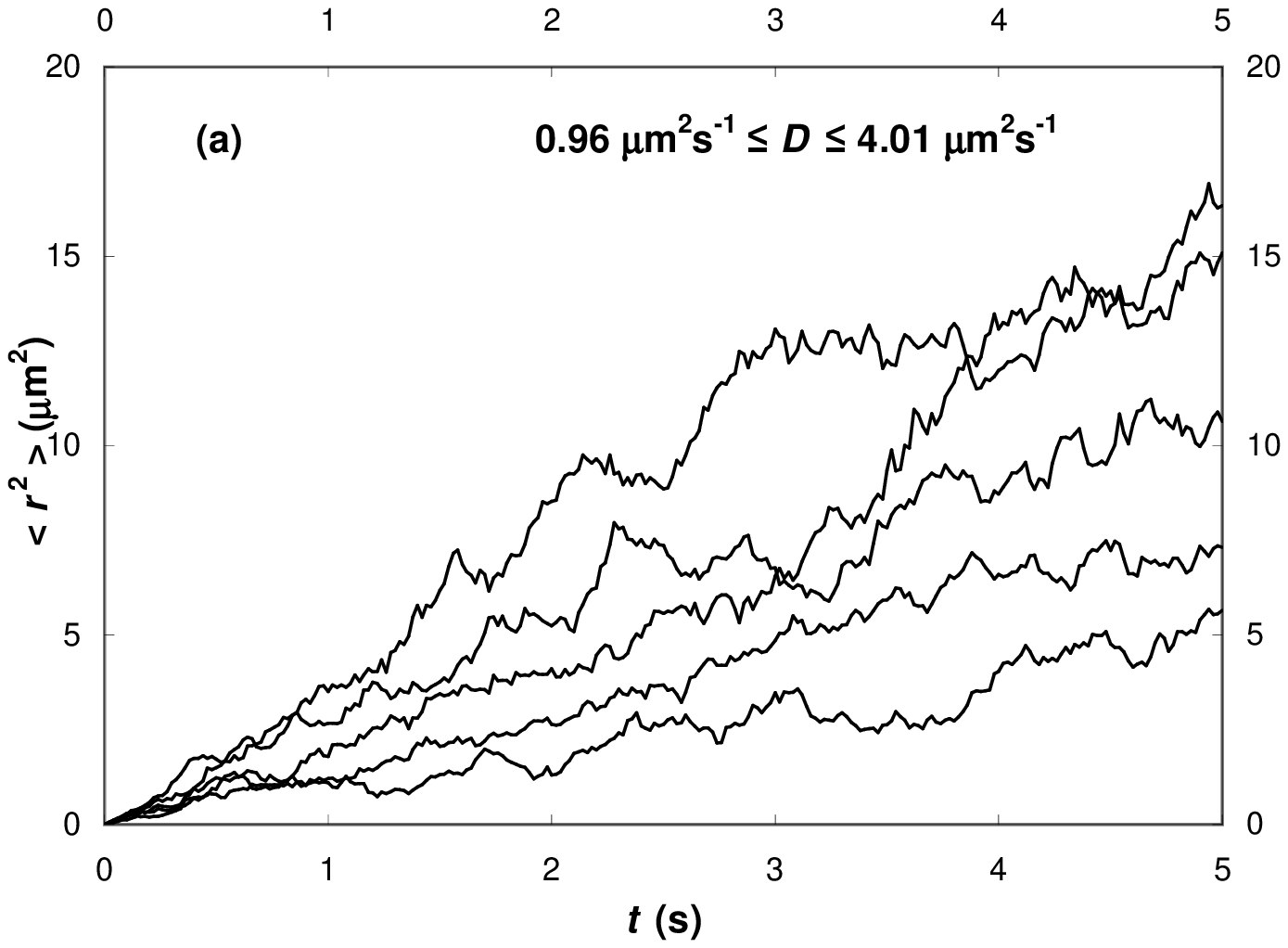}
         \includegraphics[height=5truecm,angle=0]{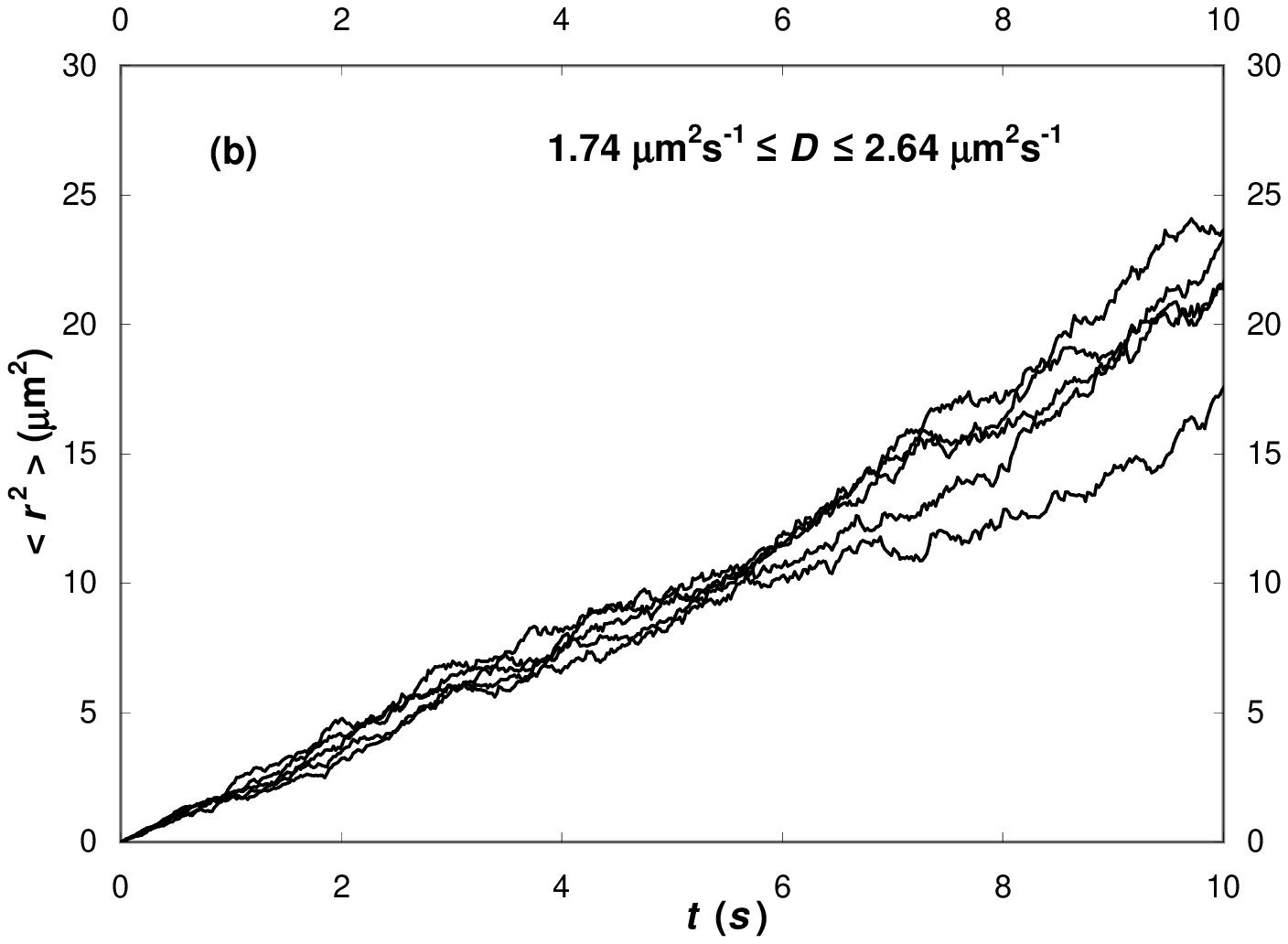}
         \\[12pt]
          \includegraphics[height=5truecm,angle=0]{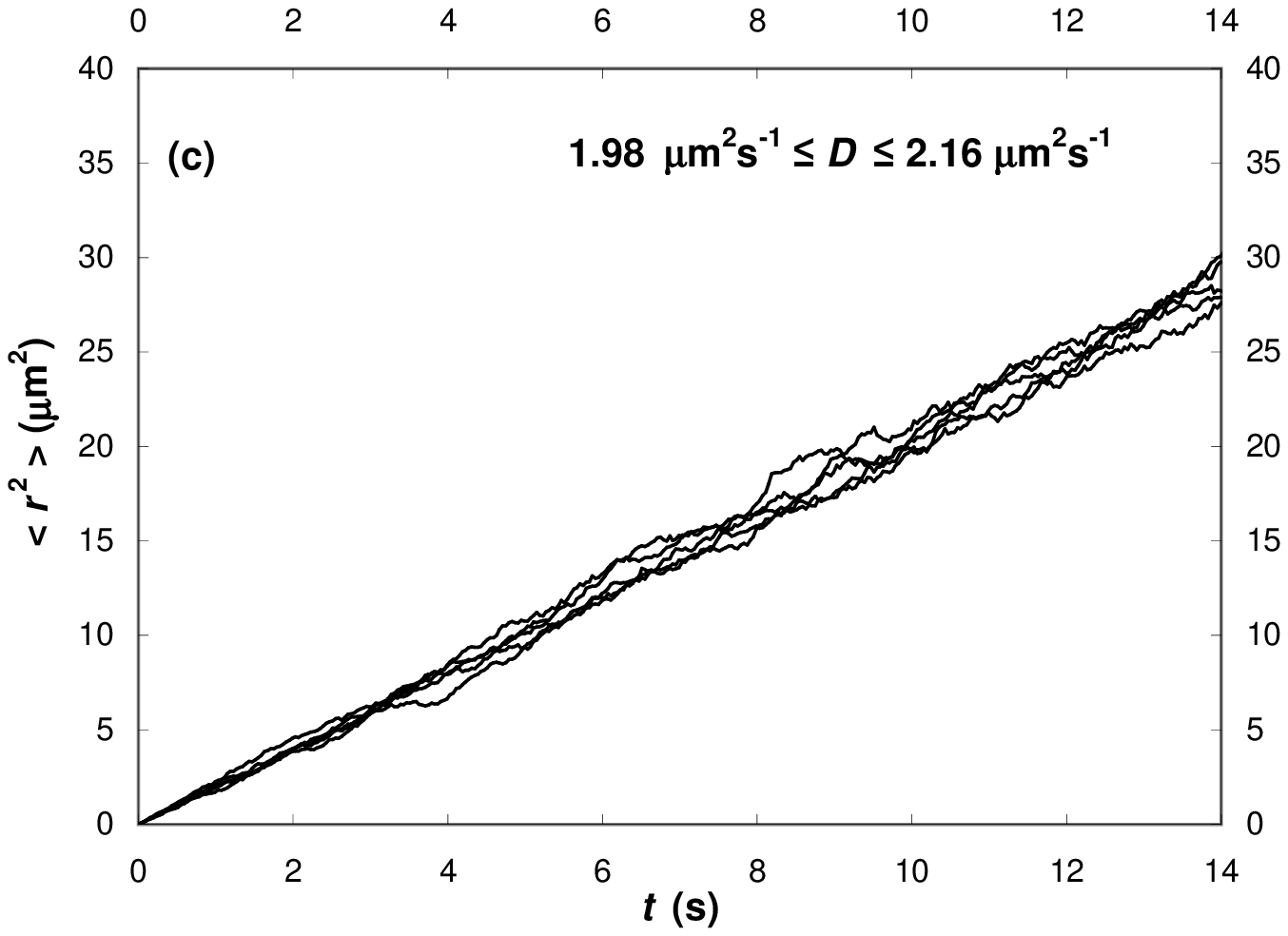}
          \includegraphics[height=5truecm,angle=0]{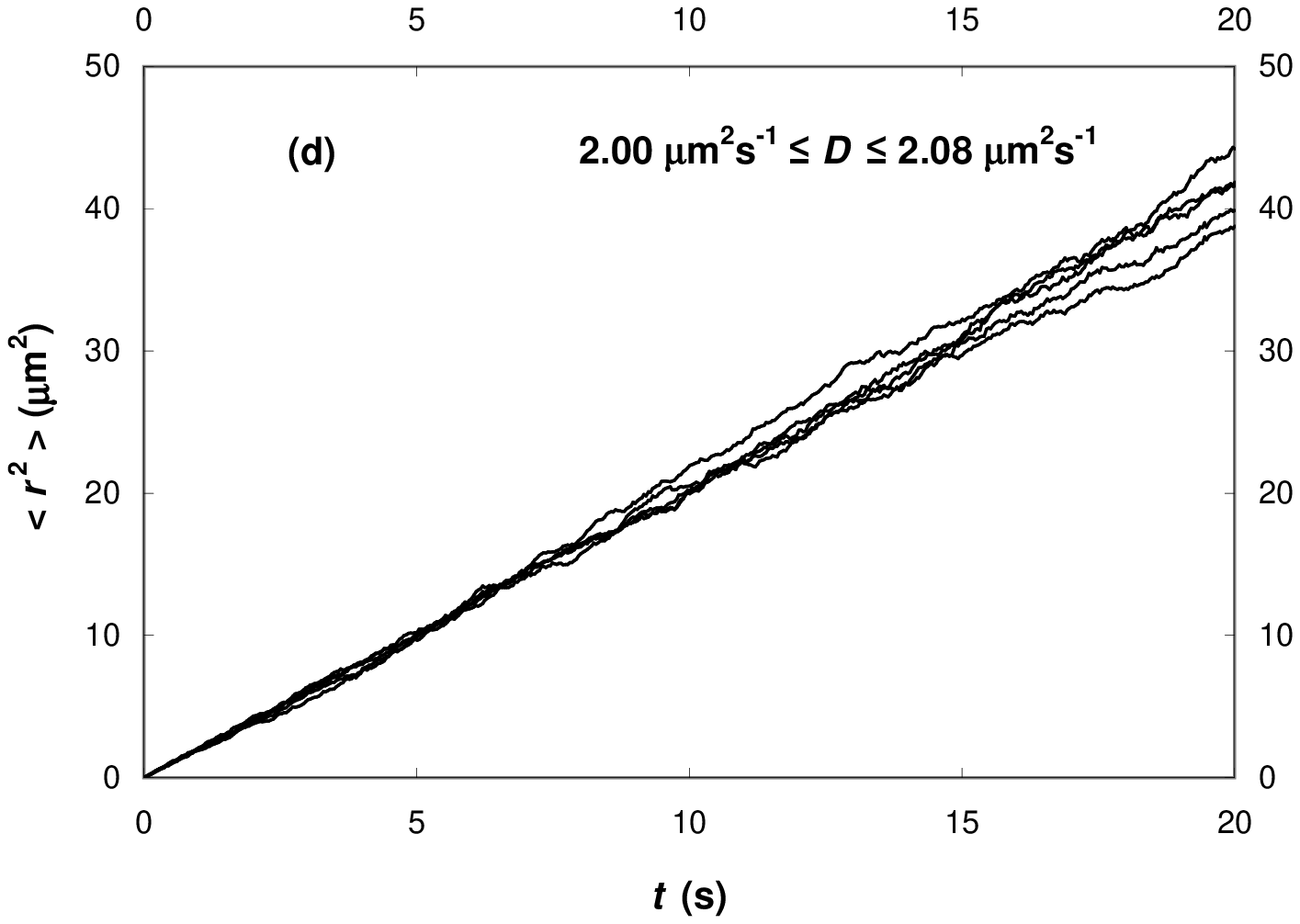}
  \end{center}
    \label{figdgzm3}
     \caption{The range of diffusion constants obtained for $n=10$(a),
$n=50$ (b), $n=160$(c) and $n=320$(d) tracked particles in
numerical simulation.}
     \end{figure}

\begin{figure}
    \begin{center}
 \includegraphics[height=5truecm,angle=0]{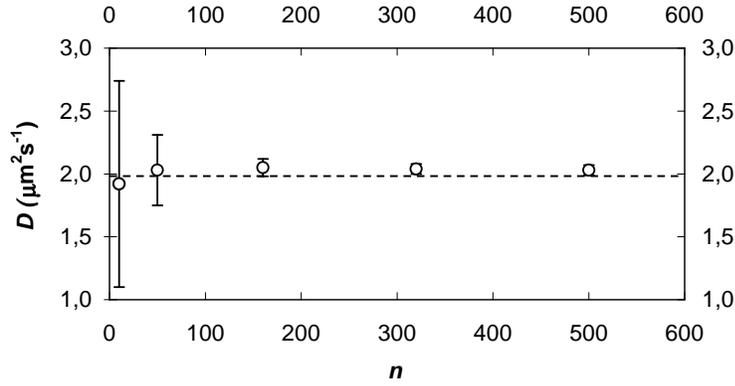}
  \end{center}
    \label{figdgzm4}
      \caption{The mean $D$ values as the function of $n$ obtained via
numerical simulation. The vertical lines represent the standard
deviation. The dotted horizontal line points the expected
theoretical value $D_{th}= 2.01 \mu m^2s^{-1}$.}
     \end{figure}

     \begin{figure}[tbh]
    \begin{center}
        \includegraphics[height=6truecm,angle=0]{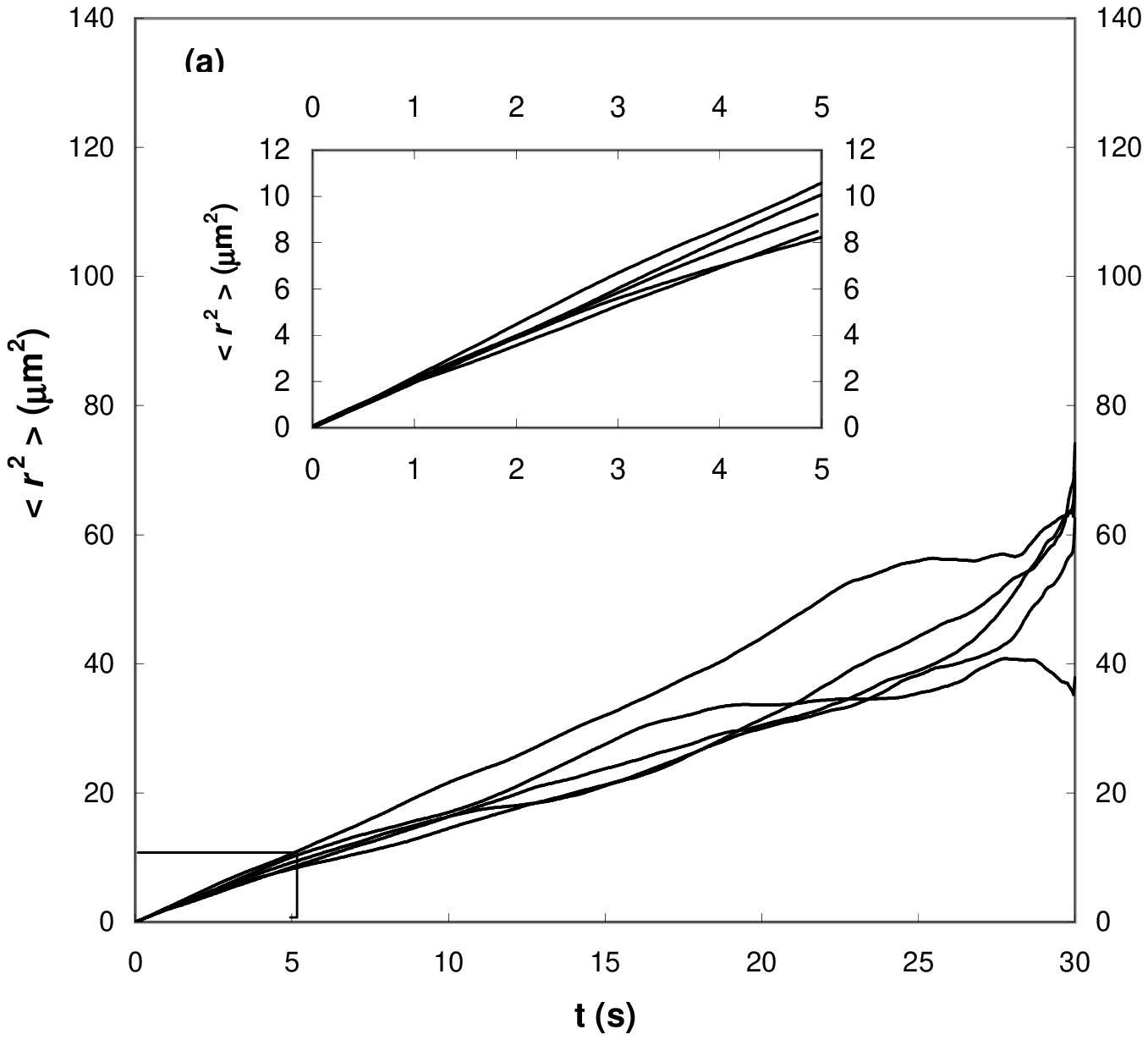}
        \includegraphics[height=6truecm,angle=0]{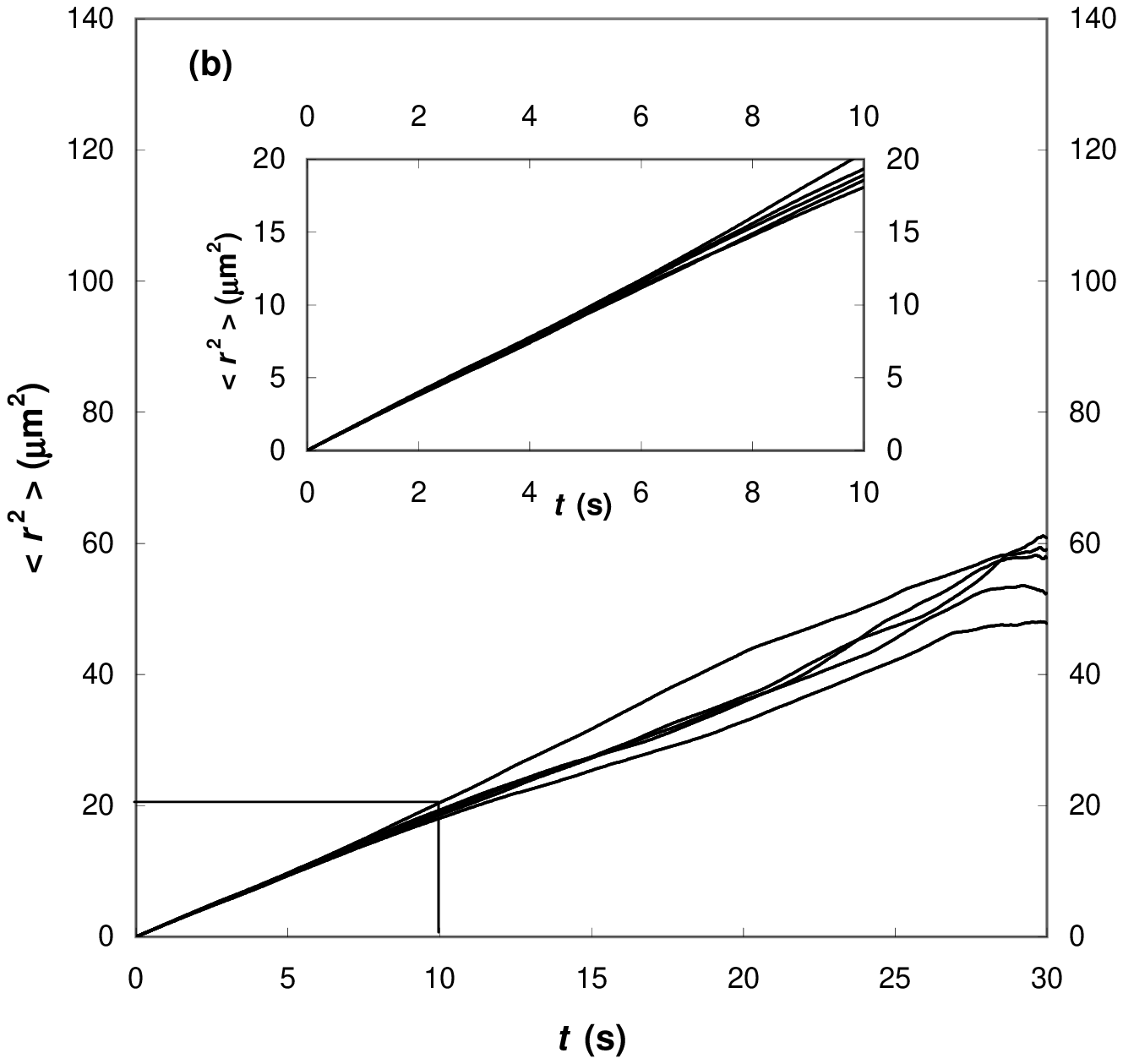}
  \end{center}
    \label{figdgzm5}
     \caption{The bunch of $\langle r^2\rangle \sim t$ for $n=10$(a)
and $n=50$(b)particles worked out with AIS. The enlarged scaling
range regime is shown separately.}
     \end{figure}

     \begin{figure}[tbh]
    \begin{center}
        \includegraphics[height=5truecm,angle=0]{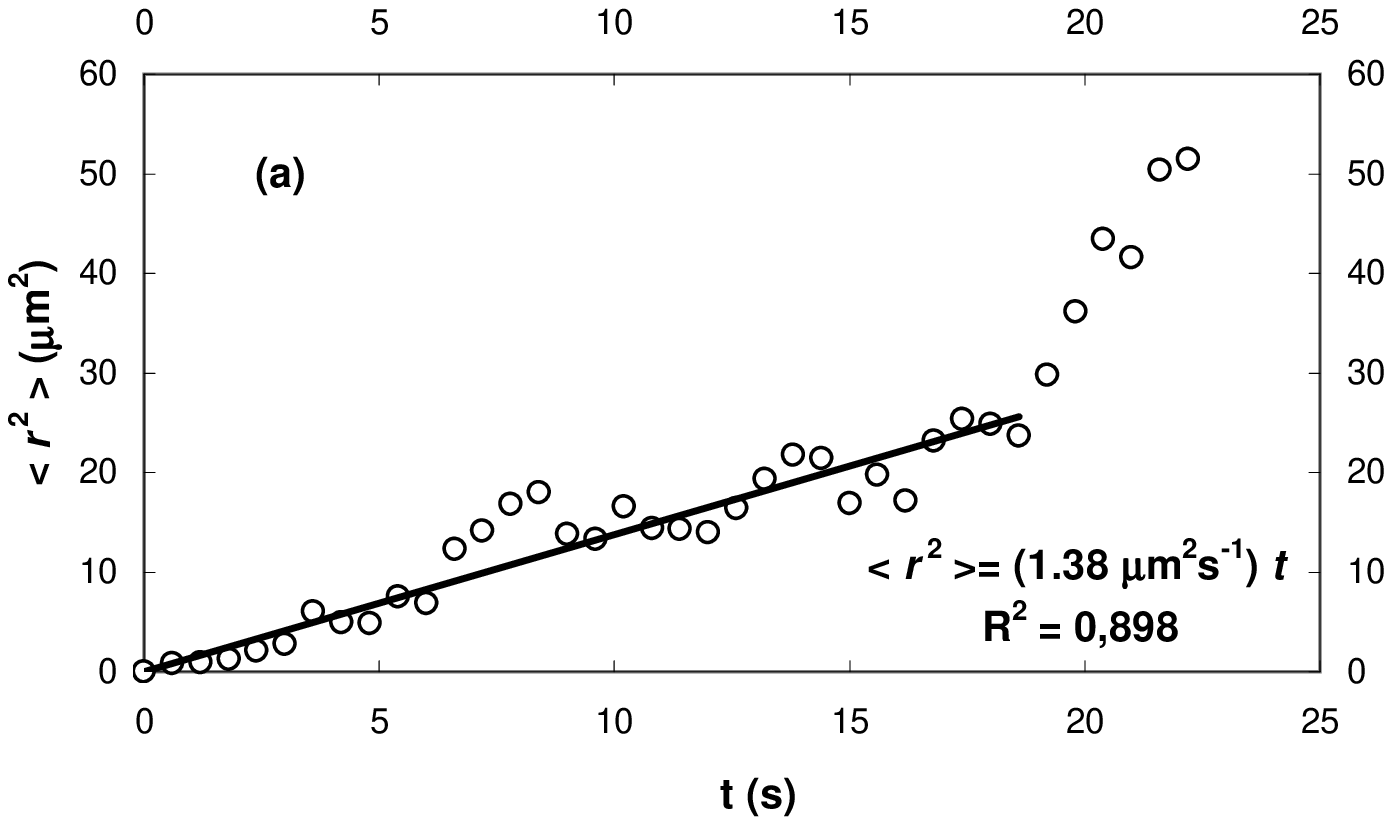}
        \\[8pt]
          \includegraphics[height=5truecm,angle=0]{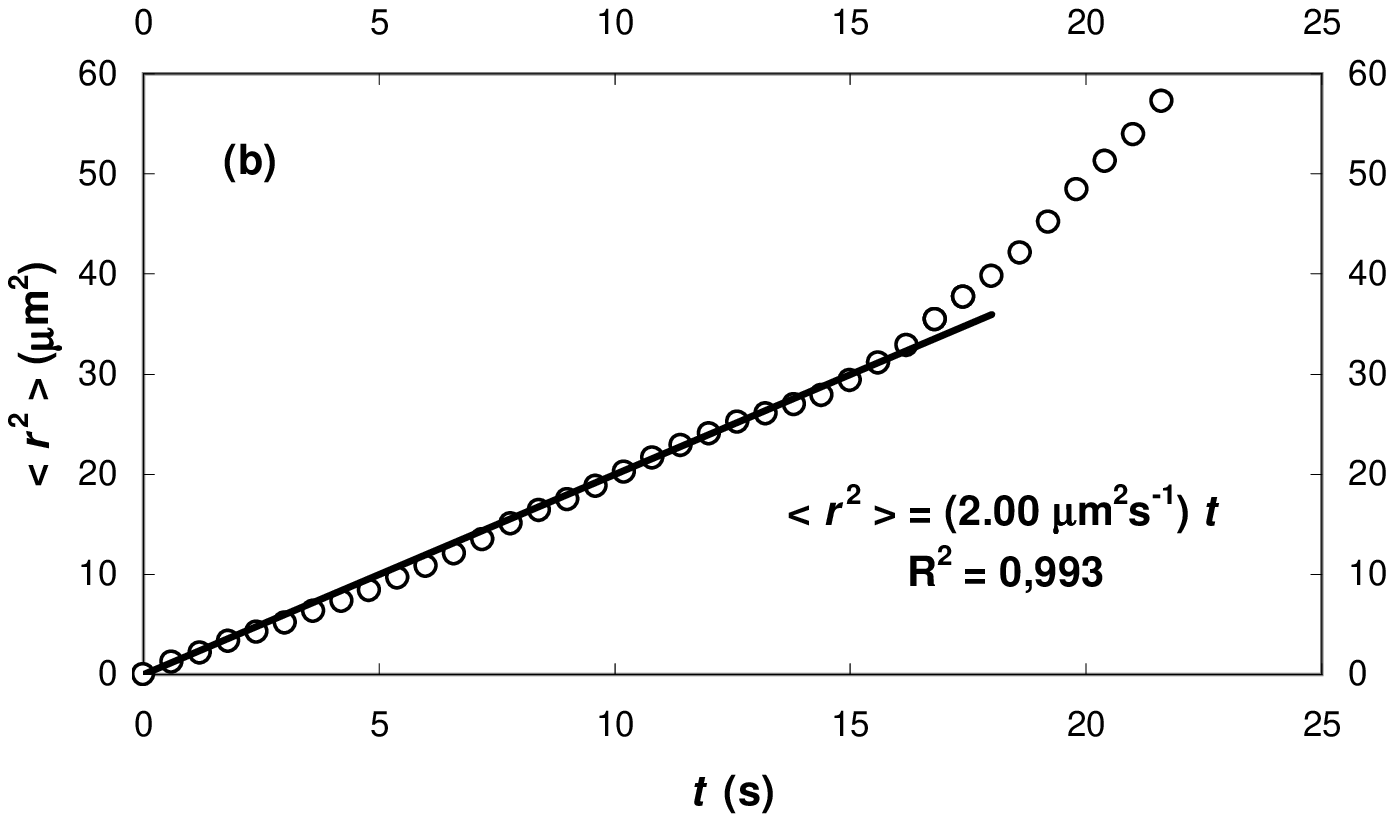}
  \end{center}
    \label{figdgzm6}
     \caption{The result of AIS procedure applied to the real
experimental data of $n=5$ latex spherical particles diffusing in
pure water (data taken from Ref.~[4]). Fig.6a represents the
"naked" results, while Fig.~6b shows the results lifted by AIS. }
     \end{figure}

  \end{document}